\documentclass[reprint,aps,pra]{revtex4-1}

\usepackage{amsmath,amsfonts,amssymb,mathrsfs,graphicx,subcaption,float}

\begin{document}
\title{Topological aspects of periodically driven non-Hermitian Su-Schrieffer-Heeger model}	
	
\author{Vivek M. Vyas}

\author{Dibyendu Roy}

\affiliation{Raman Research Institute, Bangalore 560080, India}

\begin{abstract}
 A non-Hermitian generalization of the Su-Schrieffer-Heeger model driven by a periodic external potential is investigated, and its topological features are explored. We find that the bi-orthonormal geometric phase acts as a topological index, well capturing the presence/absence of the zero modes. The model is observed to display trivial and non-trivial insulator phases and a topologically non-trivial M{\"o}bius metallic phase. The driving field amplitude is shown to be a control parameter causing topological phase transitions in this model. While the system displays zero modes in the metallic phase apart from the non-trivial insulator phase, the metallic zero modes are not robust, as the ones found in the insulating phase. We further find that zero modes' energy converges slowly to zero as a function of the number of dimers in the M{\"o}bius metallic phase compared to the non-trivial insulating phase.
\end{abstract}	
	
\maketitle

\section{Introduction}

In the last few decades, the existence and consequences of topological quantum numbers in condensed matter systems have been a subject of great interest \cite{thouless1998topological}. Unlike symmetry-based Noether conservation laws, the nature and origin of topological objects and corresponding invariants are fundamentally different. A macroscopic condensed matter system's physical state possessing a non-trivial topological invariant is, by definition, found to be immune to disorder and external perturbations \cite{bernevig2013}. There has been a lot of activity to exploit this robustness for diverse quantum applications like lasing and computation \cite{ozawa2019}.

It is a well-known fact that the time evolution of a closed quantum system is described by a Hermitian Hamiltonian, which gives rise to a unitary time evolution. Nevertheless, most quantum systems in practice interact with the external environment, and hence their physics can not always be captured by a closed system description \cite{breuer2002}. However, solving a macroscopic quantum system's dynamics while employing a general open system approach is usually daunting. In this scenario, a prudent compromise is to use an effective non-Hermitian Hamiltonian to describe the quantum system at hand, wherein the non-Hermiticity capture, in essence, the effects like dissipation arising due to the interaction with the environment \cite{vedral2003}. In any case, in reality, most quantum condensed matter systems are coupled to the measuring apparatus in some way, which in turn can give rise to non-Hermiticity in the system, as was shown in the case of the Kitaev chain recently \cite{kawabata2019,ueda2018p,Bondyopadhaya2020}.

While the topological aspects of closed quantum systems have been the main focus for a long time \cite{thouless1998topological,bernevig2013}, of late, the study of topological properties of non-Hermitian Hamiltonians has gained significant attention \cite{chiu2016,ozawa2019,ueda2018}. It is now understood that the topological structure of the non-Hermitian system is much richer and diverse than their Hermitian counterparts \cite{berry2004,yeon2019, xiong2018,nori2017}. In this context, the concepts of bulk-boundary correspondence, topological invariants, and geometric phases have been a subject of scrutiny for a while \cite{nori2017, Yao2018, kunst2018, lieu2018, kawabata2019, song2019, das2019, borgnia2020, Xiao2020, Helbig2020}. 

In condensed matter systems, the bedrock underneath the topological aspects is a discrete spatial translational symmetry in the system, giving rise to the Brillouin zone in the $k$-space \cite{bernevig2013,thouless1998topological}. The topological invariant like Chern number and the geometric phase, e.g., the Pancharatnam-Zak phase \cite{pz2019}, owe their genesis to this framework. Interestingly, there is an ingenious way of generalizing the same mathematical structure in the temporal domain by driving the system externally by a periodic potential  \cite{leon2013, dalibard2014}. Such driving can effectively add another dimension to the spatially periodic lattice; thus, a driven lattice system's topological properties are fundamentally different from the undriven case \cite{demler2010,lindner2011}. This feature is seen in several studies on these systems, and various techniques to correctly capture the topological aspects of such systems, based on the Floquet analysis, have been developed \cite{roy2017,wang2017, gong2018f, ZhouPRB2019, ZhouPan2019, HongPRB2020, ZhouPRB2020}. With the advent of precision cold atoms and quantum materials experiments, topological aspects of such a driven system have been explored in different experimental setups \cite{zoller2011,weitenberg2016, SatoPRB2019, McIver2020}.

As noted earlier, the non-Hermitian systems display a richer topological structure than the Hermitian ones; one naturally wonders what happens if such a system is periodically driven. What is the proper topological invariant capturing the topological phases of such a system, and can these phases of the system all be probed by tuning the driving potential; are some questions that immediately arise. In this work, we study the topological properties of a non-Hermitian extension of the celebrated Su-Schrieffer-Heeger (SSH) model driven by an external AC electromagnetic field and answer the above queries.

We begin by investigating a non-Hermitian version of the SSH model, possessing chiral invariance. Constructing a proper gauge-invariant geometric phase viz. bi-orthonormal generalization of Pancharatnam-Zak phase, which is acquired by a Bloch state as it traverses a circuit in the Brillouin zone, we find that the system exhibits three distinct topological phases (a) trivial insulator, (b) non-trivial insulator, and (c) M{\"o}bius metallic phase.  In the two insulator phases, the model displays a band gap in the energy spectrum's real part. The geometric phase is $0$ and $\pi$ for the trivial and non-trivial insulator phases, respectively. When the model is considered with an open boundary condition (OBC), it is found that the non-trivial insulating phase displays doubly degenerate zero modes as in the Hermitian case. In the M{\"o}bius metallic phase, we observe that the two bands of the energy spectrum's real part are merged into one, giving rise to a metallic behavior; at the same time, the state space displays M{\"o}bius strip topology as explained later. Owing to this non-trivial topology, a proper definition of the geometric phase requires the Bloch state to traverse two circuits in the Brillouin zone, and the bi-orthonormal geometric phase hence acquired, is found to be $\pi$. While the system in this metallic phase does display the existence of doubly degenerate zero modes, as one would expect, they, in turn, are found to be embedded in the band continuum and hence are not robust. The lack of robust zero modes in the M{\"o}bius metallic phase indicates that the topological notion of bulk-boundary correspondence in this model, as found in its Hermitian counterpart, is strictly speaking absent.

We further explore the scaling of zero-mode energy in the open chains as a function of the number of dimers $N$ in the non-trivial insulating and M{\"o}bius metallic phase. We find that the two phases display different scaling behavior as a function of $N$ when we correctly implement the OBC using a large $N$ limit. The energy of zero modes converges slowly to zero with $N$ in the M{\"o}bius metallic phase compared to the non-trivial insulating phase.   

\begin{figure}
	\includegraphics[scale=0.42]{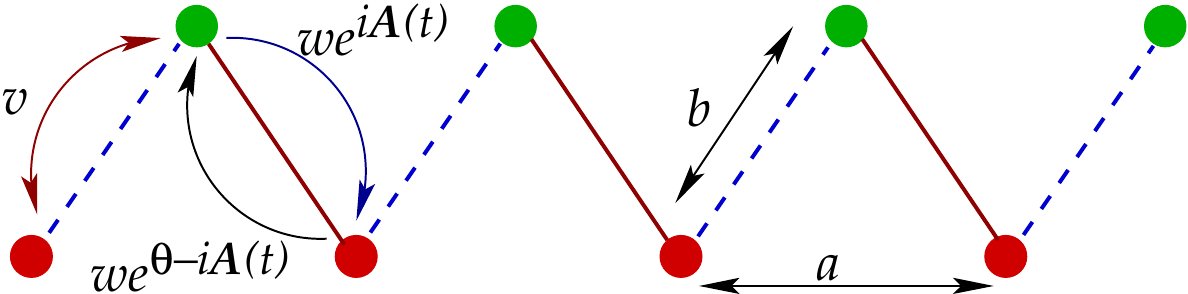}\caption{\label{schfig} A schematic diagram of the driven non-Hermitian SSH model. The red and green circles denote $A$ and $B$ sublattice sites, respectively. The lattice constant is $a$, and two sites in an unit cell are separated by distance $b$. The intracell hopping amplitude is $v$, whereas the intercell amplitudes are $w e^{i\textbf{A}(t)}$ and $w e^{\theta-i\textbf{A}(t)}$ respectively for to and fro tunnelling in the presence of a time-periodic vector potential $\textbf{A}(t)$ and a non-Hermiticity measure $\theta$. }
\end{figure}

Subsequently, we study the driven non-Hermitian SSH model, minimally coupled to an external time-periodic vector potential (see Fig.~\ref{schfig}). This model possesses two discrete translation symmetries respectively in space and time and hence allows us to solve for the Floquet-Bloch states, the spatiotemporal generalization of Bloch states. Being mainly interested in diabatic dynamics, we first consider the regime wherein the driving frequency is much larger than the model's hopping parameters. Extending the notion of the bi-orthonormal Pancharatnam-Zak geometric phase to this driven model, we find that this geometric phase indeed captures the quintessential topological features in this case as well. Upon solving the Floquet-Bloch eigenvalue problem in this large-frequency regime, we find that the system, akin to the undriven model, admits three distinct topological phases: the trivial insulator, non-trivial insulator, and M{\"o}bius metallic phase. We further support these phases' appearances by showing the presence/absence of zero-quasienergy modes in the real part of the quasienergy spectra of the driven model with an OBC using a large $N$ limit. We discover that the system's topological phase can be entirely altered by the amplitude of the driving potential, which plays as a control parameter (see Fig.~\ref{schfigP}). In the opposite limit, wherein the driving frequency is much smaller than hopping parameters, the system's dynamics essentially is adiabatic. Then topological features of the system do not get affected by external driving. Finally, we present numerical results for the driven non-Hermitian SSH model's quasienergy spectra at an intermediate frequency. These spectra show that both the non-trivial and the M{\"o}bius metallic phases can reappear when the driving's amplitude is tuned. 

\section{Non-Hermitian SSH model}

We here consider a non-Hermitian generalization of the SSH model, a one-dimensional nearest-neighbor tight-binding model comprising two sublattice sites within the unit cell (Fig.~\ref{schfig} with $\textbf{A}(t)=0$). In its generality, this model is defined in terms of four complex numbers $v_{1,2}$ and $w_{1,2}$, which parameterize the system Hamiltonian \cite{lieu2018}:
\begin{align} \nonumber
H =& \: \sum_{n=1}^{N} \left( v_1 |n,B\rangle \langle n,A| + v_2 |n,A\rangle \langle n,B|  \right)  \\ \nonumber
&+ \sum_{n=1}^{N-1} \left( w_1 |n+1,A\rangle \langle n,B| + w_2 |n,B\rangle \langle n+1,A| \right)  \\
& +  l \left( w_1 |1,A\rangle \langle N,B|  + w_2 |N,B\rangle \langle 1,A| \right). \label{HSSH}
\end{align}
The state $|n, A(B) \rangle$ represents the localized state in the $n^{th}$ unit cell with sublattice index $A(B)$. Here, the lattice constant is $a$, and the distance between the two sublattice sites $A$ and $B$ in the unit cell is given by $b$. The parameter $l$ in the above Hamiltonian is employed to ensure the validity of appropriate boundary conditions. For example, the choice $l=1$ represents the periodic boundary condition (PBC) wherein the unit cell translation operator $T(a)$ commutes with $H$: $ [ T(a), H] = 0$. Evidently, for any other choice of $l$, such discrete translation symmetry is absent. In the study of topological aspects, when $l=0$ is of particular interest, it represents the OBC \cite{bernevig2013}. For any value of parameters $v_{1,2}$, $w_{1,2}$ and $l$, this system possesses chiral symmetry, which means that the Hamiltonian $H$ anticommutes with the operator $\Sigma$, $\{ \Sigma, H \} = 0$, where:
\begin{align}
\Sigma = \sum_{n=1}^{N} \left( |n,A\rangle \langle n,A| - |n,B\rangle \langle n,B| \right).
\end{align} 
This anti-commutativity ensures that the spectrum of $H$, which is in general complex, is such that for every eigenstate $| \psi \rangle$ with energy $E$, there exists another eigenstate $\Sigma | \psi \rangle$ with energy $-E$. It further  indicates that a zero energy eigenmode is doubly degenerate. 

To understand the impact of non-Hermiticity on this model's topological aspects, we consider a special case in this work, wherein $v_1 = v_2 = v$, $w_1 = w$, $w_2 = w e^{\theta}$, for all real values of $v$, $w$, and $\theta$ with $v,w,\theta \ge 0$. In this parametrization, the value of $\theta$ can be understood to measure the divergence of the system from the Hermitian $\theta = 0$ case. 

\subsection{k-space analysis} 

When $l=1$, the existence of discrete translation symmetry allows us to convert  to $k$-space, and work with states $| k_{j}, A(B) \rangle$ (with $k_j = 2 \pi j/Na$ and $j=1,2,\dots,N$), which are the Fourier transform of localized states $|n,A(B)\rangle$. The Hamiltonian (\ref{HSSH}) in the $k$-space reads:
\begin{align}
H = \sum_{k_j}  
\begin{pmatrix}
| k_j , A \rangle, & | k_j , B \rangle
\end{pmatrix}
[ H (k_j)]
\begin{pmatrix}
\langle k_j , A| \\ \langle k_j , B|
\end{pmatrix},
\end{align} 
where the non-Hermitian matrix $H(k_j)$ reads:
\begin{align} \label{Hk} H(k_j) =
\begin{pmatrix}
0 & w e^{i k_j} + v \\ w e^{\theta - i k_j} + v  & 0
\end{pmatrix}.
\end{align}
We have assumed that the distance between the two sublattice sites is vanishing ($b=0$) and lattice constant $a=1$ for simplicity. We further set Planck constant $\hbar=1$. Notice that this Hamiltonian displays $k$-space periodicity since: $H(k=0) = H(k=2\pi)$. 

The complex energy spectrum arising from this $k$-space Hamiltonian is given by $E_{\pm}(k) = \pm E(k)$, where:
\begin{align}
 E(k) = \sqrt{ (w e^{i k} + v)(w e^{\theta - i k} + v)}, \label{enpbc}
\end{align}
which depict two complex energy bands $\pm E(k)$ in agreement with the chiral symmetry of the system. 
The corresponding left and right eigenvectors respectively are given by:
\begin{align} \label{leftvec}
\Phi_{\pm}(k) &= \frac{e^{i \lambda(k)}}{\sqrt{2}} 
\begin{bmatrix}
1, \pm \frac{E(k)}{(w e^{\theta - i k} + v)}
\end{bmatrix},\\ \label{rightvec}
\Psi_{\pm}(k) &= \frac{e^{-i \lambda(k)}}{\sqrt{2}} 
\begin{bmatrix}
1 \\ \pm \frac{E(k)}{(w e^{i k} + v)}
\end{bmatrix}.
\end{align}
Here, ${ \lambda(k)}$ stands for some arbitrary function of $k$, not necessarily respecting the periodicity in $k$. These two eigenvectors are arranged to respect the bi-orthonormality conditions: $\Phi_{\pm}(k) \cdot \Psi_{\pm}(k) = 1$ and $\Phi_{\mp}(k) \cdot \Psi_{\pm}(k) = 0$. This property of 	bi-orthonormality is a natural generalization of the orthonormality property displayed by eigenvectors of a Hermitian matrix \cite{lieu2018}. By employing this different orthonormalisation, we depart from the usual Dirac's inner product and normalization definition. The left and right Bloch eigenstates of $H$ are related to these eigenvectors respectively as:
$\langle \Phi_{\pm} (k)| = \Phi_{\pm,1}(k) \langle k ,A | + \Phi_{\pm,2}(k) \langle k ,B |$ and $| \Psi_{\pm} (k) \rangle = \Psi_{\pm,1}(k) | k ,A \rangle + \Psi_{\pm,2}(k) | k ,B \rangle$. Being the eigenstates of a non-Hermitian operator, these states do not respect the orthonormality, which follows from the Dirac inner product definition: $ \langle \Psi_{\pm} (k) | \Psi_{\mp} (k) \rangle \neq 0 $ and so on. As a result we choose to work with the bi-orthonormality conditions: $ \langle \Phi_{\pm} (k) | \Psi_{\mp} (k) \rangle = 0$, and $ \langle \Phi_{\pm} (k) | \Psi_{\pm} (k) \rangle = 1$.

 It can be checked that these Bloch eigenstates indeed obey the Bloch condition of periodicity modulo an overall phase: $T(a)| \Psi_{\pm} (k) \rangle = e^{i k}| \Psi_{\pm} (k) \rangle$. The non-trivial content of the Bloch states is contained  in the cell periodic Bloch states $|u_{\pm}(k) \rangle (\equiv T(a) |u_{\pm}(k) \rangle)$, that can be readily found from the Bloch states by applying the momentum translation operator $\mathcal{T}(-k)$, which in this case reads:
\begin{align} \label{udef}
|u_{\pm}(k) \rangle &= \Psi_{\pm,1}(k) | 0 ,A \rangle + \Psi_{\pm,2}(k) | 0 ,B \rangle, \\ \label{vdef}
\langle \tilde{u}_{\pm} (k) | &= \Phi_{\pm,1}(k) \langle 0 ,A | + \Phi_{\pm,2}(k) \langle 0 ,B |.
\end{align}  
Evidently the above right and left cell periodic Bloch states solve the eigenvalue problem for the Hamiltonian $H_{k} = \mathcal{T}(-k) H \mathcal{T}(k)$, so that $H_{k} |u_{\pm}(k) \rangle = \pm E(k) |u_{\pm}(k) \rangle$, and $\langle \tilde{u}_{\pm} (k) | H_{k} = \pm E(k) \langle \tilde{u}_{\pm} (k) |$.

\begin{figure}[t]
\begin{subfigure} {8cm}
	\includegraphics[scale=0.575]{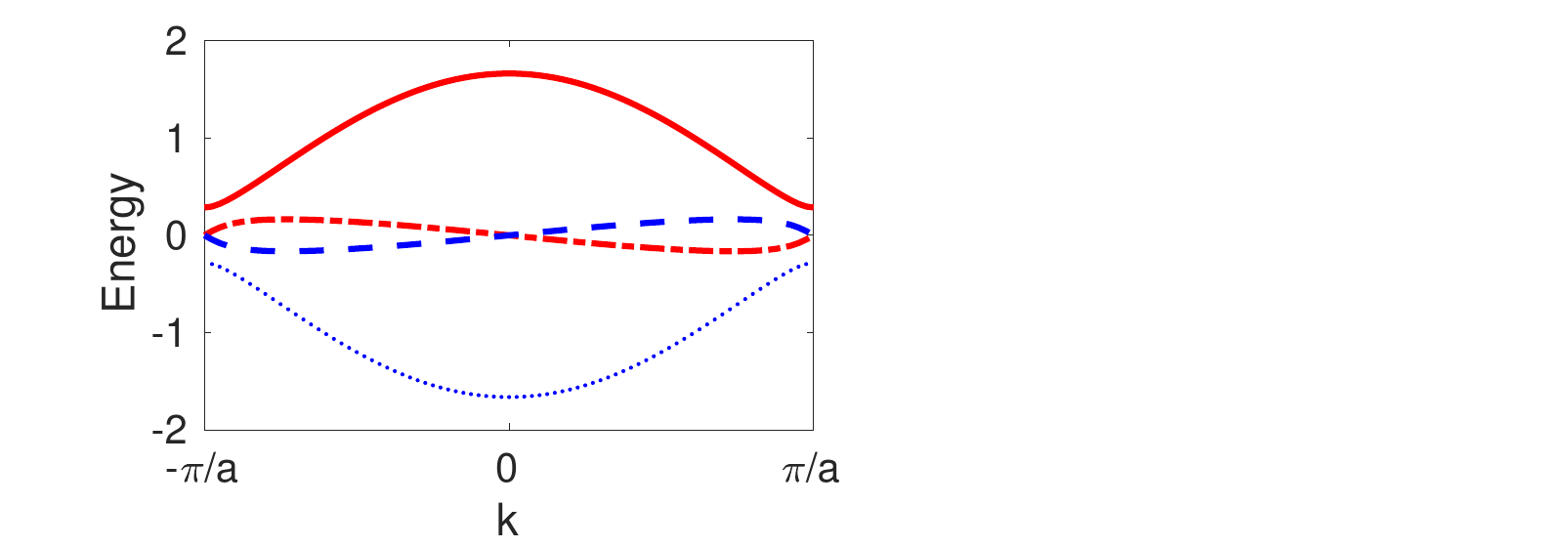}\subcaption{\label{figband1}For $v =1$, $\theta = 0.5$, and $w = -0.1 + e^{-\theta}$.  Topologically the band structure is like a cylinder with the curves $\pm E(k)$ defining the edges.}
\end{subfigure}	
\begin{subfigure} {8cm} 
	\includegraphics[scale=0.575]{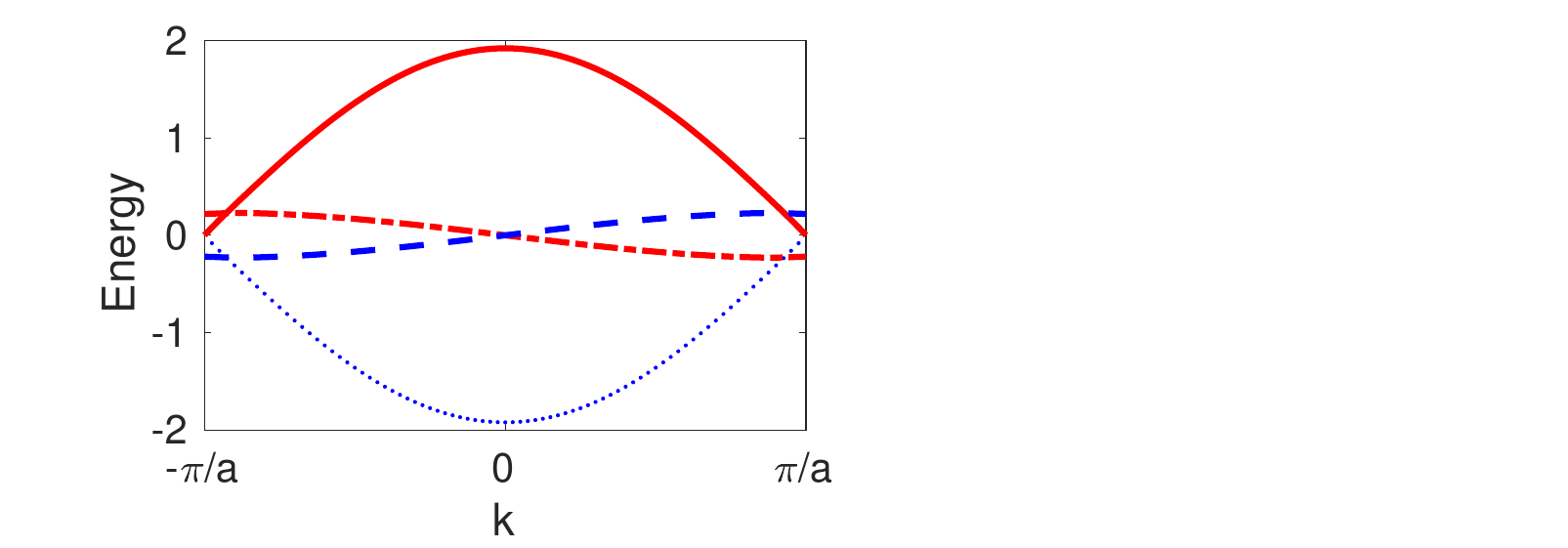}\subcaption{\label{figband2}For $v =1$, $\theta = 0.5$, and $w = 0.1 + e^{-\theta}$. Topologically the band structure is like a M{\"o}bius strip with the curve $E(k)$ defining the edge.}
\end{subfigure}	
\caption{Plot of real and imaginary parts of energy eigenvalues $E_{\pm}(k)$ as a function of $k$. Here, the red continuous and dot-dash curves respectively denote the real and imaginary part of the upper band's energy; whereas the blue dotted and dashed curves respectively indicate the real and imaginary part of the lower band's energy.}	
\end{figure}

The fact that the matrix $H(k)$ is periodic in $k$-space may lead one to naively believe that the energy spectrum and the average $\langle \tilde{u}_{\pm} (k)| \mathcal{O}| u_{\pm}(k) \rangle$ of some generic operator $\mathcal{O}$ respecting PBC, is always a periodic function of $k$, and hence return to its initial value after $2 \pi$ circuit in the $k$-space. However, this is generally not true and happens only in the region $\kappa > 1$ or $\kappa < e^{- \theta}$, where $\kappa = w/v$. In this region, the energy $E(k)$ is an analytic function of $k$, and the band structure of the system is generically given by Fig.~(\ref{figband1}). While the energy is in general complex, we can infer that the non-Hermitian system behaves as a band insulator owing to a gap in the energy spectrum's real part. Further, the eigenvectors $|\Psi_{\pm}(k + 2 \pi)|^2 = |\Psi_{\pm}(k)|^2$ and $|\Phi_{\pm}(k + 2 \pi)|^2 = |\Phi_{\pm}(k)|^2$ in this region. Therefore, the band structure as well as the state space of the system have a cylinder topology when $\kappa > 1$ or $\kappa < e^{- \theta}$. 

From the pioneering work of Pancharatnam \cite{pancha1956,mukunda1993,pz2019}, we now understand that the notion of geometric phase can be purely defined kinematically in terms of a cyclic overlap of states. In the case of the lattice model, the geometric phase referred to as Pancharatnam-Zak phase $\nu$, can be defined corresponding to a given band in terms of (Dirac normalized) cell periodic Bloch states $|u_j \rangle \equiv | u(k_j) \rangle$:
\begin{align*}
	\Delta = \langle u_0 | u_N \rangle \langle u_N | u_{N-1} \rangle \cdots
	\langle u_2 | u_{1} \rangle \langle u_1 | u_{0} \rangle,
\end{align*} 
with $\nu = \text{Arg} \:\Delta$. It is evident that $\nu$ is phase acquired by the Bloch state as it completes a circuit in the Brillouin zone. 

While the above definition of the geometric phase properly holds in the case of Hermitian models, it is not satisfactory for the present non-Hermitian case. The main reason for this is that the Bloch states living in different bands are not orthogonal regarding Dirac's definition of the inner product. 

Given this scenario, we here generalize the above definition of Pancharatnam-Zak phase to bi-orthonormal setup, which is given by $\gamma$ where $\gamma = \text{Arg} \: D$ with:
\begin{align*}
	D = \langle \tilde{u}_0 | u_N \rangle \langle \tilde{u}_N | u_{N-1} \rangle \cdots
	\langle \tilde{u}_2 | u_{1} \rangle \langle \tilde{u}_1 | u_{0} \rangle.
\end{align*} 
In the continuum limit, this phase for both the bands takes the form:
\begin{align} \label{gp1} \nonumber
	\gamma_{\pm}(2\pi) &= \text{Arg}\: (\langle \tilde{u}_{\pm}(0)| u_{\pm}(2 \pi) \rangle) \\ &+ i \int_{0}^{2 \pi} dk \:   \langle \tilde{u}_{\pm}(k) | \partial_{k} | u_{\pm}(k) \rangle.
\end{align}
By a careful choice of function $\lambda(k)$ (which amounts to fixing a gauge), the argument term can be made to vanish, in which case, this Pancharatnam-Zak phase becomes the so called complex geometric phase studied in the literature \cite{garrison1988,lieu2018}. It must be mentioned that unlike the complex geometric phase, this geometric phase is a proper geometric construct, is gauge invariant (which implies its insensitivity to any choice of $\lambda(k)$) and hence immune to any gauge redefinitions of the type: $| u_{\pm}(k) \rangle \rightarrow e^{i \Lambda(k)}| u_{\pm}(k) \rangle$, $\langle \tilde{u}_{\pm}(k) | = e^{-i \Lambda(k)} \langle \tilde{u}_{\pm}(k) |$, for some function $\Lambda(k)$. 

The above gauge-invariant definition of the bi-orthonormal geometric phase allows us to extend the results obtained in Ref.~\cite{lieu2018} for the complex geometric phase. It is known that owing to the chiral symmetry, the complex geometric phase is real and is quantized in the units of $\pi$. On evaluating (\ref{gp1}) using (\ref{udef}) and (\ref{vdef}), we immediately find that:
\begin{align*}
\gamma_{\pm}(2 \pi) &= \pi, \quad \text{when} \quad \kappa > 1, \\
& = 0, \quad \text{when} \quad \kappa < e^{-\theta}.
\end{align*}
	
In the parameter region $e^{- \theta}<\kappa < 1$, we find that the energy eigenvalue $E(k)$ as a function of $k$ is not analytic/single valued, but rather becomes \emph{double valued}, that has been studied since a while in various systems \cite{berry2004, xiong2018, nori2017}.  As a result we observe that the two limits do not agree: $E(k \rightarrow \pi) \neq E(k \rightarrow -\pi)$. We readily see that in such a case: $E_{-}(k + 2 \pi) = E_{+}(k)$, and $E_{-}(k + 4 \pi) = E_{-}(k)$. This fact is neatly brought out in the Fig. (\ref{figband2}). If we were to follow the trajectory of the complex energy in this figure starting from the lower band, then it is evident that after one circuit in $k$-space, we reach the upper band. It is only after yet another circuit that we return to the initial starting point. This is a clear evidence of the double valued nature of the energy eigenvalues.  Since the energy eigenvalues change from lower to upper band after completing a circuit in $k$-space, it also follows that the eigenvectors must follow the same route: $|\Psi_{\pm}(k + 2 \pi)|^2 = |\Psi_{\mp}(k)|^2$ and $|\Psi_{\pm}(k + 4 \pi)|^2 = |\Psi_{\pm}(k)|^2$. These results indicate that the state space of right eigenvectors is like a M{\"o}bius strip, with the edge of the M{\"o}bius strip being the states $\Psi_{\pm}(k)$. 

Physically, the two bands are not distinct or disjoint, but in fact are connected/merged into one another in a sense that the system can be continuously changed to go from one band to another. Thus, the system in this regime behaves as a metal, and the band indices $\pm 1$ are redundant since the bands are merged together. Owing to the M{\"o}bius strip nature, we need to extend the notion of circuit in $k$-space as a loop from $k=0$ to $k=4 \pi$. In fact after one $2 \pi$ circuit, we have $\Psi_{\mp}(2 \pi) = \Psi_{\pm}(0)$, so that $\Phi_{\pm}(0) \cdot \Psi_{\pm}(2\pi)  = 0$. As a result, we can not define geometric phase using relation (\ref{gp1}) since the initial and final states are orthogonal. Instead, we are required to consider $4 \pi$ circuit geometric phase:
\begin{align} \label{gp2} \nonumber
\gamma (4\pi) &= \text{Arg}\: (\langle \tilde{u}_{\pm}(0)| u_{\pm}(4 \pi) \rangle) \\ &+ i \int_{0}^{4 \pi} dk \:   \langle \tilde{u}_{\pm}(k) | \partial_{k} | u_{\pm}(k) \rangle.
\end{align}
We find from explicit calculation $\gamma (4\pi)=\pi$ through out this region of $e^{- \theta}<\kappa < 1$. This result indicates that although the system is a gapless metal but it is in a topologically non-trivial phase in this parameter region. 

\subsection{Real space analysis}

It is a well known fact that in the presence of PBC, when the Hermitian SSH model displays non-trivial value of geometric phase, the same in the presence of OBC displays doubly degenerate zero-energy modes protected from the band gap. Such a result is often referred to as the bulk-boundary correspondence, and in the non-Hermitian lattice systems this concept has off late generated a lot of interest \cite{kunst2018,song2019,borgnia2020,nori2017}. 

In order to investigate whether such a correspondence exist in the model at hand, let us consider the real space Hamiltonian (\ref{HSSH}) without assuming anything about the form of $l$. The right eigenvalue problem for this Hamiltonian is expressible as:
\begin{align}
H | \psi_{\zeta} \rangle = E_{\zeta} | \psi_{\zeta} \rangle,
\end{align}
where the eigenvalues and eigenvectors are indexed by quantum number $\zeta$. This eigenvalue problem can also be alternatively viewed as a matrix eigenvalue problem:
\begin{align}
\sum_{s'=A,B;n'=1}^{N} \langle n,s | H | n',s' \rangle \langle n',s' | \psi_{\zeta} \rangle = E_{\zeta} \langle n,s | \psi_{\zeta} \rangle,
\end{align}
where we use the completeness property of localized states $| n,s \rangle$, with site index $n=1,2,\cdots,N$, and sublattice index $s=A,B$. Alternatively this can be written as:
\begin{widetext}
\begin{align} 
\begin{bmatrix}
0 & v & 0 & 0 & \cdots & 0 &l w \\
v & 0 & w e^{\theta} & 0 & \cdots & 0 & 0\\
0 & w & 0 & v & \cdots & 0 & 0 \\
\vdots & \vdots & \vdots & \vdots & \vdots & \vdots & \vdots\\
l w e^{\theta} & 0 & \cdots & \cdots & 0 & v & 0
\end{bmatrix}
\begin{bmatrix}
\langle 1,A | \psi_{\zeta} \rangle \\
\langle 1,B | \psi_{\zeta} \rangle\\
\langle 2,A | \psi_{\zeta} \rangle\\
\vdots \\
\langle N,B | \psi_{\zeta} \rangle
\end{bmatrix}
= E_{\zeta}
\begin{bmatrix}
\langle 1,A | \psi_{\zeta} \rangle \\
\langle 1,B | \psi_{\zeta} \rangle\\
\langle 2,A | \psi_{\zeta} \rangle\\
\vdots \\
\langle N,B | \psi_{\zeta} \rangle
\end{bmatrix}.
\end{align}
\end{widetext}
Evidently the trace of $H$ is $0$, which can be understood to arise from the chiral symmetry of the model. The determinant of $H$ can be readily found from the above matrix, and it reads:
\begin{align} \label{DetH}
&&\text{Det} \; H = (-1)^N \left( {v}^{{N}} +(-1)^{N-1} l e^{{N \theta}} {w}^{{N}} \right) \nonumber \\&&\times \left( {v}^{{N}} + (-1)^{N-1}l {w}^{{N}} \right).    
\end{align}
This expression, which holds for any value of $l$ and $N$, is obtained from mathematical induction and can be straight away checked using symbolic computation. 

Let us consider the Hermitian case ($\theta=0$) for a moment. It is well known fact that the determinant of a matrix is a product of its eigenvalues. In the PBC case, we analytically know that the system admits two eigenvalue bands. In the OBC case with $l=0$, it is known that the system admits apart from the two bands, two midgap states with energy close to zero. In order to pull out the behaviour of these energy levels near zero energy, we study the ratio of two determinants respectively, of the system with OBC made up of $N+1$ dimers, and PBC with $N$ dimers:
\begin{align} \label{ratio}
r = \frac{\text{Det} H \lvert_{OBC, N+1}}{\text{Det} H \lvert_{PBC, N}}.
\end{align} 	
The ratio $r$ is thus: 
\begin{align}
r = E_{+,0} E_{-,0},
\end{align}	
where $E_{\pm,0}$ are the two energy levels closest to zero. It is evident that when the system does not admit any zero-energy modes, this ratio will be equal to $-1$; and if zero-energy modes exists then the ratio must tend to $0$ as $N \rightarrow \infty$.  It is clear that when $\kappa < 1$, the power $\kappa^{N}$ converges to $0$ as $N \rightarrow \infty$. Thus the ratio $r \rightarrow -1$ when $N \rightarrow \infty$ while $\kappa < 1$. On the other hand when $\kappa > 1$ the object $\kappa^{N}$ becomes larger and larger as $N$ grows, thus the ratio can now be written as an exponential $r \simeq  -\kappa^{-2N}$. This is in agreement with the earlier known results \cite{bernevig2013}.

Extending the above treatment to non-Hermitian system ($\theta \ne 0$), considering $N$ to be odd and $v=1$ for specificity but without losing generality, we immediately see from (\ref{DetH}) and (\ref{ratio}) that the ratio reads:
\begin{align}
r = -\frac{1}{(1 + \kappa^{N})(1 + (\kappa e^{\theta})^{N})}.
\end{align}
As observed earlier, when $\kappa > 1$, this ratio displays exponential drop $r \simeq -e^{-N\theta}\kappa^{-2N}$ in the large $N$ limit. When $e^{-\theta} < \kappa < 1$, we find that the ratio reads:
\begin{align}
r \simeq -e^{-N\theta}\kappa^{-N}.
\end{align}  
The ratio in this case also shows an exponential drop albeit slower than the insulator case viz. $\simeq -e^{-N\theta}\kappa^{-2N}$. Interestingly we see that while the zero modes exist in both non-trivial insulating and metallic phases, their  scaling as a function of system size has a characteristic difference: the drop is sharper in the insulating phase in comparison to the metallic phase.

\begin{figure}
\includegraphics[width=7.5cm,height=9cm]{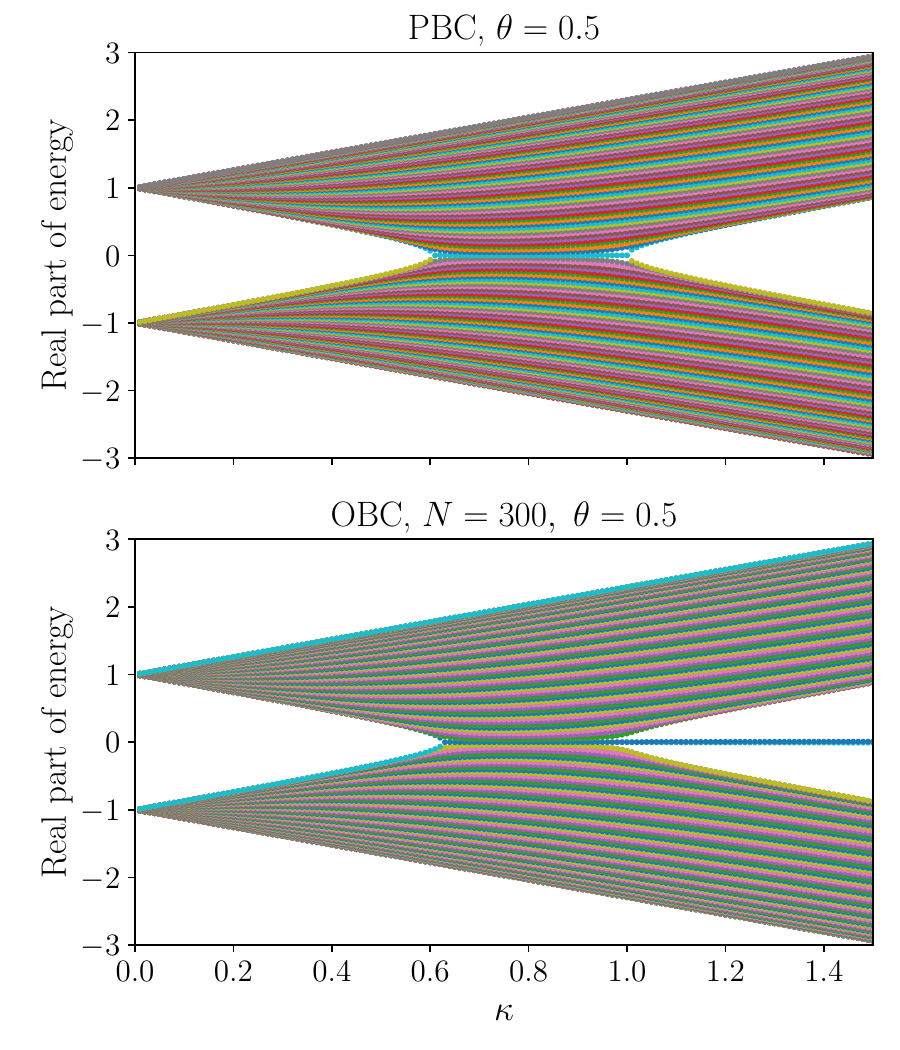}
\caption{Plots showing an agreement between analytically obtained energy spectrum with PBC (top) and numerically obtained energy spectrum for $N=300$ with OBC (bottom). The real part of energy is plotted as a function of $\kappa$ for $\theta = 0.5$. Apart from the zero modes, the two spectra closely match. \label{specplot1}}
\end{figure}

We notice that the scaling of zero-mode energies $r$ as a function of $N$ is explicitly dependent upon the non-Hermiticity parameter $\theta$. As a result, the drop in the zero-mode energies is not uniform, and systems with different $\theta$ values show different scaling behavior. This dependence of zero-mode energy scaling, being an explicit function of $\theta$, is an unpleasant feature. We wonder if we can get a better understanding of this scaling phenomenon. Often in several condensed matter systems, the technique of large $N$ expansion is employed to understand better the underlying physics  \cite{imbo1985,jevicki1981}. In particular, this technique has found immense applications in the study of quantum phase transitions \cite{sachdev2011} and is known to be a non-perturbative technique.  

Now we implement the OBC in our problem by employing the large $N$ technique, and we consider $l N = \text{constant}~(=1)$ in which case the determinant at hand reads:
\begin{align}
&&\text{Det} \; H = (-1)^N\left( {v}^{N} + (-1)^{N-1}\frac{1}{N} e^{{N \theta}} {w}^{{N}} \right) \nonumber \\&& \left( {v}^{{N}} + (-1)^{N-1} \frac{1}{N} {w}^{{N}} \right),
\end{align} 
and the ratio now reads (for non-zero but small $\theta$ and $w$ around 1):
\begin{align}
r \simeq -\frac{(1 - \frac{1}{N}\kappa^{N})(1 - \frac{1}{N}(\kappa e^{\theta})^{N})}{(1 + \kappa^{N})(1 + (\kappa e^{\theta})^{N})}.
\end{align}
When $\kappa > 1$, which corresponds to the {non-trivial insulator phase}, both the exponentials $\kappa^{N}$ and $(\kappa e^{\theta})^{N}$ grow indefinitely as $N \rightarrow \infty$, and so the ratio can be approximated as:
\begin{align}
r \simeq -\frac{(- \frac{1}{N}\kappa^{N})(- \frac{1}{N}(\kappa e^{\theta})^{N})}{(  \kappa^{N})(  (\kappa e^{\theta})^{N})} = -\frac{1}{N^2}.
\end{align}
This shows that the zero-mode energies are real and they fall as $\frac{1}{N}$, that is, $E_{+,0}=\frac{1}{N}$ and $E_{-,0}=-\frac{1}{N}$. We note that unlike the usual OBC condition $l=0$, here the fall of zero-mode energies is primarily independent of the non-Hermiticity parameter $\theta$ or $\kappa$.  

When the system is in the metallic phase, which means $e^{-\theta} < \kappa < 1$,  we find that the exponential $\kappa^N$ converges to $0$ as $N \rightarrow \infty$, while the exponential $(\kappa e^{\theta})^N$ grows indefinitely. As a result, the ratio can be written as:
\begin{align}
r \simeq  -\frac{(1 - \frac{1}{N}(\kappa e^{\theta})^{N})}{(1 + (\kappa e^{\theta})^{N})} = \frac{1}{N}.
\end{align} 
The above implies that the zero-mode energies are imaginary, and they fall as $\frac{1}{\sqrt{N}}$, that is, $E_{+,0}=\frac{i}{\sqrt{N}}$ and $E_{-,0}=-\frac{i}{\sqrt{N}}$. Thus, the zero-mode energies in the metallic phase converge to zero slower as $\frac{1}{\sqrt{N}}$ rather than the $\frac{1}{N}$ in the insulator phase, but the once again the scaling property is independent of the Hamiltonian parameters. The imaginary energy of zero modes as well as their slower convergence to zero with system size in the metallic phase further indicate the non-robustness of these modes there.

The different scaling of zero-mode energy as a function of $N$ in the two topological phases displays a kind of universality since the scaling property is sensitive only to the topology and not the actual values of the Hamiltonian parameters.  Apparently this scaling behaviour is not specific to the choice of Hamiltonian parameters as $v_1 = v_2 = v$, $w_1 = w$, $w_2 = w e^{\theta}$, but rather is a feature displayed by general non-Hermitian chiral Hamiltonian (\ref{HSSH}) for any complex choices of $v_{1,2}$ and $w_{1,2}$, as is shown in the Appendix.

The above discussion shows that the implementation of OBC employing a large  $N$ limit captures the topological features of the model, the zero-energy modes, which begin right from the metallic M{\"o}bius phase itself and extend to the non-trivial insulator phase as well: $\kappa > e^{-\theta}$.  Nevertheless, the physical significance of zero modes, which are embedded in the gapless continuum of energy levels in the metallic phase, is diminished in the M{\"o}bius region. The physical reason is that there is no gap to distinguish these modes from the other infinitesimally close by levels. As a result, these modes are not robust compared to the zero modes in the non-trivial insulator region. The plots in Fig. (\ref{specplot1}) of the real part of the spectrum for both PBC and OBC show a close agreement, modulo the zero modes, validating the above-presented analysis. The above treatment also shows that the topological properties of the system with OBC manifested as zero-energy modes are well captured by the non-trivial value of the bi-orthonormal geometric phase in both insulator and metallic phases. The topological phase structure of this model can be summarized as:

\begin{enumerate}
	\item[A.]\underline{Trivial insulator phase}: $\kappa < e^{-\theta}$, wherein $\gamma(2 \pi)=0$. The bands have cylinder topology.
	\item[B.]\underline{Non-trivial insulator phase}: $\kappa > 1$, wherein $\gamma(2 \pi)=\pi$. The bands have cylinder topology, and robust zero modes exist.
	\item[C.]\underline{M{\"o}bius metallic phase}: $e^{-\theta} < \kappa <1 $, $\gamma(2 \pi)$ is not defined, but $\gamma(4 \pi) = \pi$. The bands are merged into M{\"o}bius strip topology, and zero modes exist but are not robust.
\end{enumerate}	
The above analysis further indicates that there is an absence of bulk-boundary correspondence in the non-Hermitian SSH model, when we consider the M{\"o}bius metallic phase as topologically non-trivial, but the zero modes there are not robust.

\section{Driven Non-Hermitian SSH model}
	
We now discuss a scenario wherein the non-Hermitian SSH model, defined by (\ref{HSSH}), is driven by an external electromagnetic field represented by a vector potential $\textbf{A}(t)$. The minimal Hamiltonian that depicts such a driven model can be written as \cite{leon2013}: 	
\begin{widetext}
\begin{align} \nonumber
H_{D}(t) &= \: \sum_{n=1}^{N} \left( v |n,B\rangle \langle n,A| + v  |n,A\rangle \langle n,B| \right)
+ \sum_{n=1}^{N-1} \left( w e^{i \textbf{A}(t)} |n+1,A\rangle \langle n,B| + w e^{\theta-i \textbf{A}(t)} |n,B\rangle \langle n+1,A| \right) \\ 
& +  l \left( w e^{i \textbf{A}(t)} |1,A\rangle \langle N,B|  + w e^{\theta-i \textbf{A}(t)} |N,B\rangle \langle 1,A| \right).
\label{DHSSH}
\end{align}
\end{widetext}
In this discussion, we shall assume that the external electromagnetic field is sinusoidally varying in time: $\textbf{A}(t) = A_{0} \sin \Omega t$, with a period $T = \frac{2 \pi}{\Omega}$, where $A_0$ is the amplitude of the vector potential and $\Omega$ is its frequency. This provides us with an interesting physical setup wherein the system possesses two independent discrete translational symmetries (assuming the PBC):
\begin{align*}
[H_{D}(t), T(a)] = 0, \quad \text{and} \quad [ H_{D}(t), U(T) ] = 0. 
\end{align*}
Here, $U(T)$ represents the discrete time translation operator for time period $T$: $| \varphi (t + T) \rangle = U(T) | \varphi (t) \rangle$, where $| \varphi (t) \rangle$ is some generic state. 

Owing to the fact that the system at hand is periodically driven in time, we wish to find the analogues of Bloch eigenstates in this case, albeit which solve the time-dependent Schrodinger equation as also respect the Bloch property spatially and temporally when the PBC is imposed in both real space and time. The states obeying the Bloch property in the temporal domain are well studied in literature and are referred to as Floquet states \cite{shirley1965,dalibard2014,leon2013}. Let us denote the Floquet states obeying the spatial Bloch property as $| Y_{\alpha} (k,t) \rangle$ which are defined as:
\begin{align}
&\left( H_{D}(t) - i \partial_{t}\right) | Y_{\alpha} (k,t) \rangle = 0, \\
& T(a) | Y_{\alpha} (k,t) \rangle = e^{i k} | Y_{\alpha} (k,t) \rangle, \\ & U(T) | Y_{\alpha} (k,t) \rangle = e^{-i \varepsilon_{\alpha}(k) T} | Y_{\alpha} (k,t) \rangle.
\end{align} 
Here, the quasienergy $\varepsilon_{\alpha}(k)$ is the temporal analogue of wave vector $k$, and $\alpha$ is the band index. Akin to the wavevector, the quasienergy is also ambiguous up to an addition of integer multiples of $\Omega$ \cite{shirley1965}. 

As observed earlier in the undriven SSH model, the geometric phase and the topological aspects associated with the system are captured by the cell periodic Bloch states. Here, we can correspondingly define the cell periodic Floquet-Bloch states $| u_{\alpha} (k,t) \rangle$ as $| Y_{\alpha} (k,t) \rangle = e^{- i \varepsilon_{\alpha}(k) t} \mathcal{T}(k) | u_{\alpha} (k,t) \rangle$, which solve the eigenvalue problem:
\begin{align} \label{fbproblem}
\left( \mathcal{T}(-k) H_{D}(t) \mathcal{T}(k) - i \partial_{t}\right) | u_{\alpha} (k,t) \rangle = \varepsilon_{\alpha}(k) | u_{\alpha} (k,t) \rangle,
\end{align} 
and obey: $T(a) | u_{\alpha} (k,t) \rangle = | u_{\alpha} (k,t) \rangle$, and $U(T) | u_{\alpha} (k,t) \rangle = | u_{\alpha} (k,t) \rangle$.

This eigenvalue problem can be well tackled by going over to the temporal domain Fourier space, by invoking the completeness properties of the Fourier modes $| m \rangle \equiv e^{-im\Omega t}$, and using the Sambe inner product: $\langle \langle A,m | B,m' \rangle \rangle = \frac{1}{T} \int_{0}^{T} dt\: e^{i (m-m')\Omega t} \langle A | B \rangle$ \cite{shirley1965}. This allows us to write the above eigenvalue problem as:
\begin{align}
\sum_{m=-\infty}^{\infty}
\langle \langle m' | \mathcal{H}_{k}(t) | m \rangle \rangle \langle \langle m | u_{\alpha}(k,t) \rangle \rangle = \varepsilon_{\alpha}(k) \langle \langle m' | u_{\alpha}(k,t) \rangle \rangle,
\end{align}
where the Floquet-Bloch operator $\mathcal{H}_{k}(t) \equiv \left( \mathcal{T}(-k) H_{D}(t) \mathcal{T}(k) - i \partial_{t}\right)$. This problem can be further simplified by going over to the $k$-space, in which case the matrix elements of the Floquet-Bloch operator $\mathcal{H}_{k}(t)$ can be straight away evaluated and expressed using the Bessel functions $J_{m}(x)$ to read:
\begin{align} \nonumber
&\langle \langle m' | \mathcal{H}_{k}(t) | m \rangle \rangle = 
\begin{pmatrix}
|0,A \rangle, & |0,B \rangle 
\end{pmatrix}
[\mathcal{H}_{m',m}(k)] \begin{pmatrix}
\langle 0,A | \\ \langle 0,B | 
\end{pmatrix},
\end{align}
where
\begin{align}
& [\mathcal{H}_{m',m}(k)] = \nonumber\\
&\begin{pmatrix}
- m' \Omega \delta_{m',m} & v + w e^{i k} J_{m-m'}(A_0) \\
v + w e^{-i k + \theta} J_{m'-m}(A_0) & - m' \Omega \delta_{m',m}
\end{pmatrix}. \label{fbmatrix}
\end{align}
So by going over to the spatial and temporal Fourier domain, the problem of finding the quasienergy spectrum for the Floquet-Bloch problem (\ref{fbproblem}) is now reduced to diagonalizing the infinite dimensional matrix $[\mathcal{H}_{m',m}(k)]$. 

In general, in the absence of any other symmetries, analytical diagonalization of $[\mathcal{H}_{m',m}(k)]$ for arbitrary values of parameters is an arduous task. As a result, we are forced to work with a judicious approximation, which renders this diagonalization possible and maximally captures the accurate quasienergy spectrum. 

{\it High-frequency driving:} We note that the matrix $[\mathcal{H}_{m',m}(k)]$ is dominated by the diagonal elements in the Fourier space so long as $\Omega >> v,w,e^{\theta},J_{0}(A_0)$. To the leading order in this limit, we can approximate this matrix as being diagonal, neglecting the off-diagonal terms in the Fourier space \cite{leon2013}:
\begin{align}
[\mathcal{H}_{m',m}(k)] \simeq - m' \Omega \delta_{m',m} + \mathcal{H}_{0}(k),
\end{align}
where
\begin{align}
\mathcal{H}_{0}(k) = 
\begin{pmatrix}
0 & v + w e^{i k} J_{0}(A_0) \\
v + w e^{-i k + \theta} J_{0}(A_0) & 0
\end{pmatrix}. \label{effH}
\end{align}    
This shows that the dynamics of each Fourier block is essentially governed by the same non-trivial matrix $\mathcal{H}_{0}(k)$ while only the diagonal part changes albeit inconsequentially since the quasienergy is ambiguous upto addition of term $m'\Omega$. As a result, the essential dynamics of the system in this regime is captured by the non-Hermitian matrix $\mathcal{H}_{0}(k)$. 

Remarkably, we see that the structure of the matrix $\mathcal{H}_{0}(k)$ is identical to that of $H(k)$, encountered in (\ref{Hk}), which governs the dynamics of the undriven model. The matrix $\mathcal{H}_{0}(k)$ is indeed $H(k)$, albeit with a redefinition $w \rightarrow \tilde{w} = w J_{0}(A_0)$. With this identification, we have solved the eigenvalue problem for $\mathcal{H}_{0}(k)$, to find that the quasienergy band structure is:   
\begin{align} \label{quasiband}
\varepsilon_{\pm} (k)= \pm \sqrt{ (\tilde{w} e^{i k} + v)(\tilde{w} e^{\theta - i k} + v)}.
\end{align}
Owing to the similarity with the energy band spectrum $E_{\pm}(k)$ of undriven non-Hermitian SSH model, we can immediately infer that the quasienergy spectrum $\varepsilon_{\pm}(k)$ also realizes distinct phases depending upon the value of $\tilde{\kappa} = \tilde{w}/v$. In particular, the quasienergy band merger takes place when $e^{-\theta} < \tilde{\kappa} <1 $, and the system becomes M{\"o}bius metallic in this region, whereas in the other regions the quasienergy spectrum displays a band gap and behaves as an insulator (see Fig. (\ref{schfig})). It is interesting to note that the parameter $\tilde{\kappa}$ is a function of driving potential amplitude $A_0$ through the Bessel function $J_{0}(A_0)$, which is an oscillating function of $A_0$.  Due to the non-monotonic dependence of $\tilde{\kappa}$ on $A_0$, the system makes back and forth transitions  between the metallic and insulator phases as $A_0$ is increased monotonically. The last feature is neatly brought out in the Figs.~(\ref{quasiplot1}) and (\ref{quasiplot2}).   		
		
\begin{figure}[H]
\includegraphics[width=9.5cm]{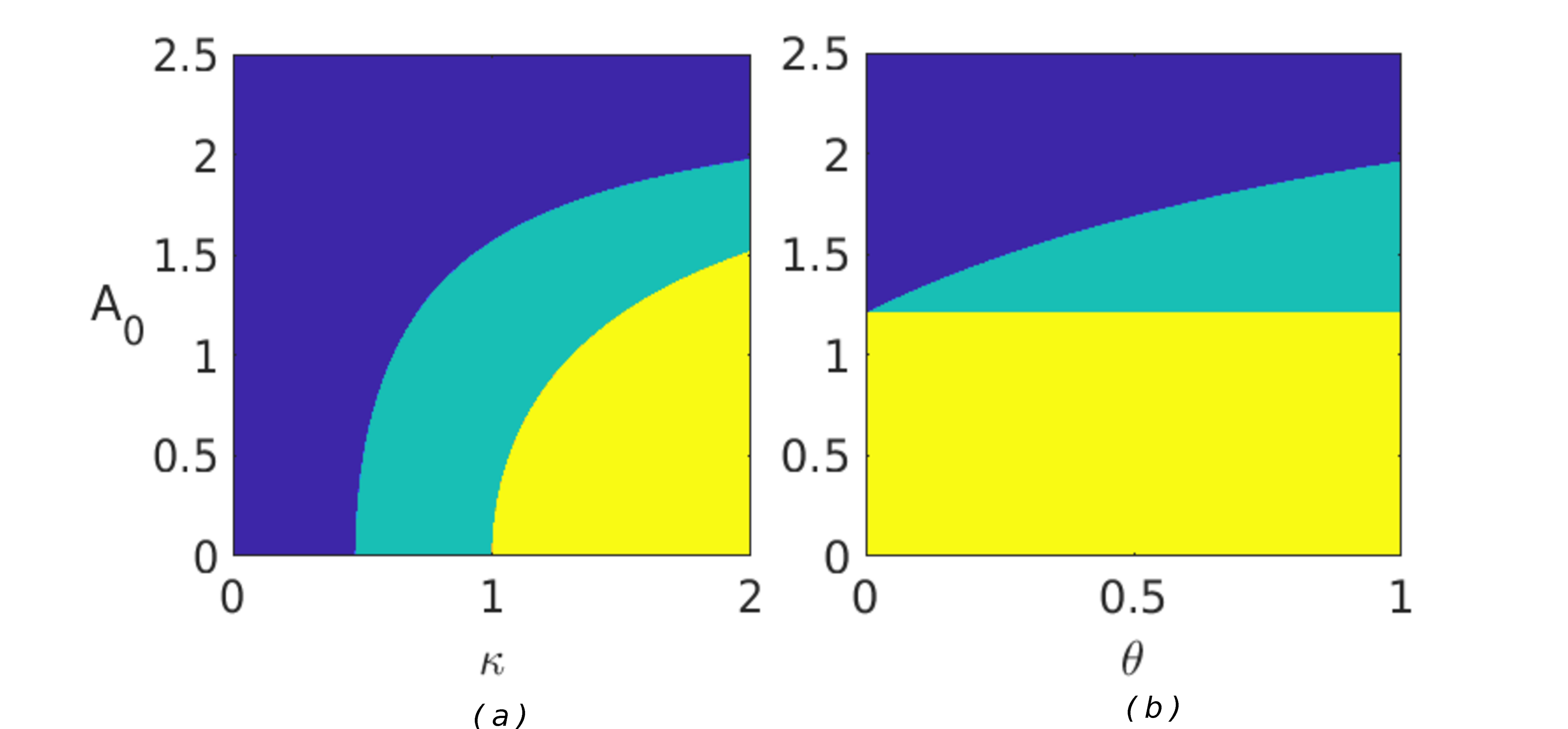}
\caption{\label{schfigP} Geometric phase $\gamma$ in a driven non-Hermitian SSH model as a function of vector potential's amplitude  $(A_0)$ and ratio $\kappa~(\equiv w/v)$ for non-Hermiticity parameter $\theta=0.75$ (a), and the same as a function of $A_0$ and $\theta$ for $\kappa=1.5$ (b). Here the blue region represents the trivial insulating phase ($\gamma(2\pi) = 0$), while the yellow region indicates the non-trivial insulating phase ($\gamma(2\pi) = \pi$). The green region denotes the M{\"o}bius metallic phase ($\gamma(4\pi) = \pi$).}
\end{figure}

\begin{figure}[h]
\includegraphics[width=8cm,height=10cm]{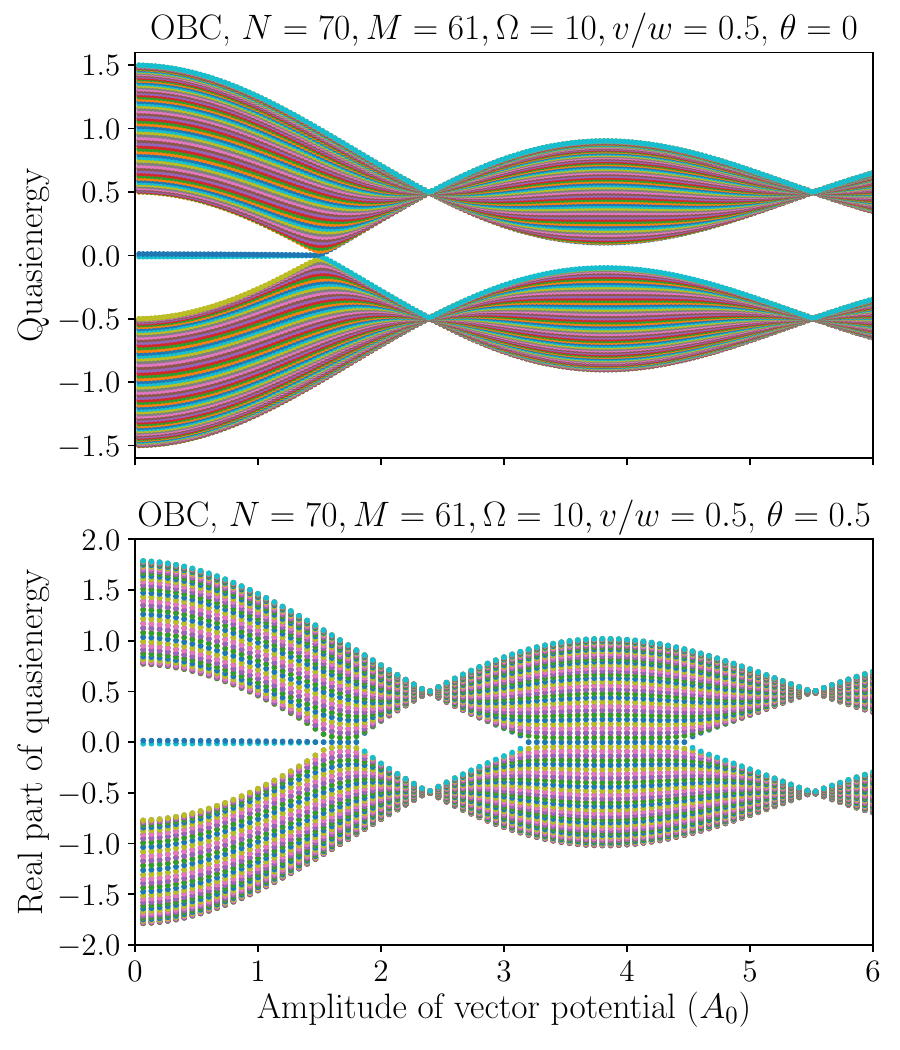}
\caption{Quasienergy spectrum versus amplitude of vector potential $(A_0)$ of a driven Hermitian SSH model (top) and a driven non-Hermitian SSH model (bottom) at high-frequency driving and with OBC. The M{\"o}bius strip  phase appears and reappears only in the driven non-Hermitian model. All parameters are shown at the headings. \label{quasiplot1}}
\end{figure}

\begin{figure}[h]
\includegraphics[width=8cm,height=10cm]{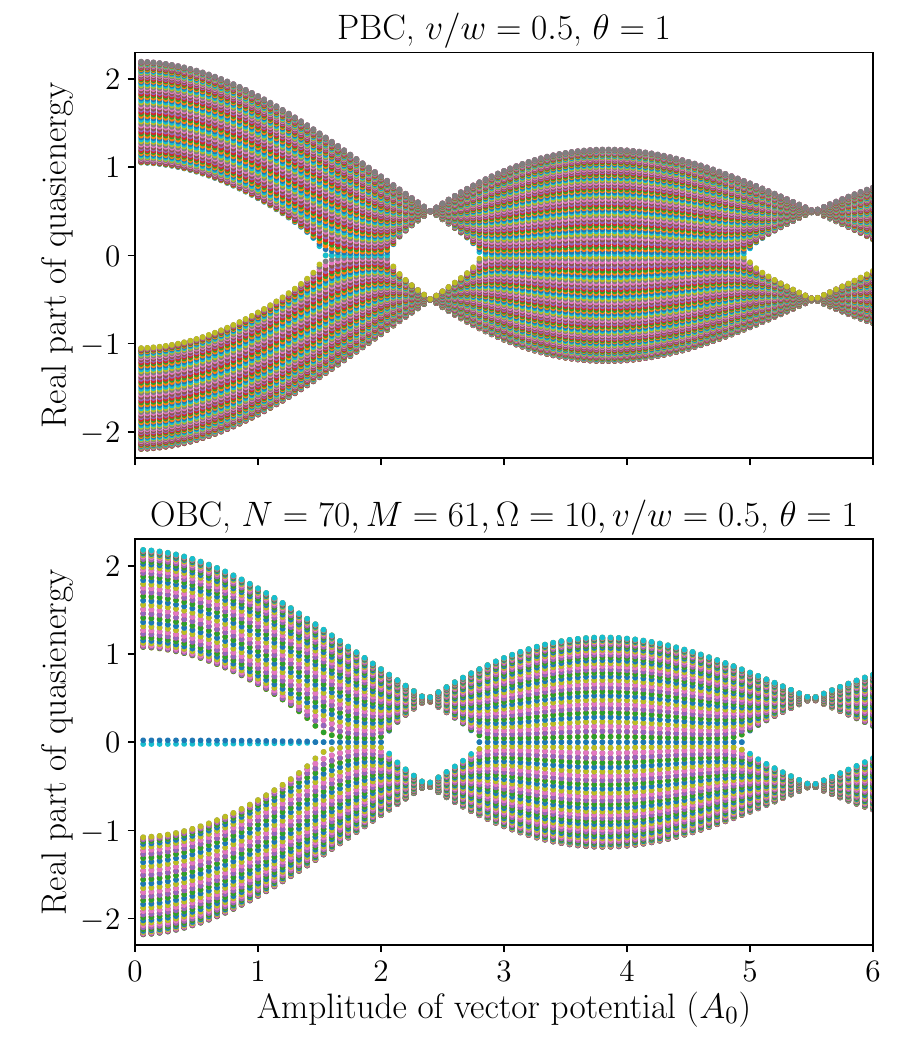}
\caption{Plots showing an agreement between analytically obtained quasienergy spectrum with PBC (top) and numerically obtained quasienergy spectrum with OBC (bottom). The real part of quasienergy is plotted as a function of amplitude of vector potential $(A_0)$ for a driven non-Hermitian SSH model at high-frequency driving. Apart from the zero modes for the spectrum with OBC, the two spectra closely match. All parameters are shown at the headings. \label{quasiplot2}}
\end{figure}

The left and right eigenstates of $\mathcal{H}_{0}(k)$ are also respectively the same as $\Phi_{\pm}$ and $\Psi_{\pm}$ in (\ref{leftvec}) and (\ref{rightvec}), with the replacement $w \rightarrow \tilde{w}$. This allows us to apply the definition of the bi-orthonormal geometric phase (\ref{gp1}) in this driven case as well, which reads:
\begin{align} \nonumber
\gamma_{\pm}(2\pi) &= \text{Arg}\: (\Phi_{\pm}(0)\cdot\Psi_{\pm}(2\pi)) \\& + i \int_{0}^{2 \pi} dk \: \Phi_{\pm}(k) \cdot \partial_{k} \Psi_{\pm}(k).
\end{align} 
We note that the Pancharatnam-Zak phase is also quantized in the units of $\pi$ in the insulator phases of the driven non-Hermitian SSH model. Upon explicit calculation, we find that the ratio $\tilde{\kappa}$ separates the trivial and non-trivial insulating phase of the driven model. When $\tilde{\kappa} > 1$, the system is in a non-trivial insulating phase characterized by $\gamma_{\pm}(2\pi) = \pi$. For $0 < \tilde{\kappa} < e^{-\theta}$, the system behaves as a trivial band insulator, wherein $\gamma_{\pm}(2\pi) = 0$. In the metallic phase, we deduce that the two circuit Pancharatnam-Zak phase $\gamma_{\pm}(4\pi)$ is the proper geometric invariant with the value $\pi$. In Fig.~(\ref{schfigP}), we plot the geometric phase as a function of $A_0$, $\theta$ and $\kappa$, bringing out the phase diagram of the driven non-Hermitian system.   

The above analysis of the driven non-Hermitian SSH model was performed while committing to PBC. Nevertheless, we wonder if the non-trivial insulating phase has a manifestation in zero-quasienergy eigenmodes in the OBC case of this driven model. To investigate this possibility, we consider the Hamiltonian in (\ref{DHSSH}), where we correctly implement the OBC by employing a large $N$ limit as $l=1/N$. Akin to the real space analysis of undriven SSH model, we now discuss the Floquet eigenvalue problem of the above Hamiltonian in real space:
\begin{align}
&\left(H_{D}(t) - i \partial_{t} \right) | \psi_{\mu} \rangle = \varepsilon_{\mu} | \psi_{\mu} \rangle.
\end{align} 
Using the completeness property of the Fourier modes $|m \rangle$ and the localized states $|n,s \rangle$, and employing the Sambe inner product, this Floquet problem reads:
\begin{widetext}
\begin{align}
\sum_{m',n',s'} \langle \langle m | \langle n,s | \left(H_{D}(t) - i \partial_{t} \right)  |n',s' \rangle | m' \rangle \rangle  \langle \langle m' | \langle n',s' | \psi_{\mu} \rangle \rangle = \varepsilon_{\mu} \langle \langle m | \langle n,s | \psi_{\mu} \rangle \rangle.
\end{align} 
\end{widetext}
Similar to the real space analysis of undriven model, the object $\langle \langle m | \langle n,s | \left(H_{D}(t) - i \partial_{t} \right)  |n',s' \rangle | m' \rangle \rangle$ can be expressed in the form of a matrix, keeping only finite number of Fourier modes (call $M$), and then be numerically diagonalized to yield the quasienergy spectrum. 

In Fig.~(\ref{quasiplot1}), we compare quasienergy spectrum obtained using the above procedure for the driven Hermitian $(\theta=0)$ and non-Hermitian $(\theta \ne 0)$ SSH model. While the quasienergy spectrum is real for the Hermitian model, it  is complex for the non-Hermitian model. Thus, we plot the real part of the quasienergy spectrum for the driven non-Hermitian SSH model. The M{\"o}bius strip  phase appears and reappears only in the driven non-Hermitian model, and it is absent in the driven Hermitian SSH model. In Fig.~(\ref{quasiplot2}), we then compare the above quasienergy spectrum with the one obtained analytically from approximate description of (\ref{quasiband}) using PBC, considering a lattice of $N=70$ and number of Fourier modes $M=61$. Apart from the zero modes for the spectrum obtained with OBC, we observe a good agreement between the spectra in the two cases. The phase diagrams in Fig.~(\ref{schfigP}) show that the topological features of the OBC spectra found in Figs.~(\ref{quasiplot1},\ref{quasiplot2}) are well captured by the geometric phase relations obtained using PBC earlier. 

{\it Low-frequency driving:} In the above discussion, it is assumed that the driving frequency $\Omega$ is much larger than the other parameters in the model. In the opposite limit, if the driving frequency $\Omega \ll v,w$, then it implies that the rate of change of Hamiltonian $\partial_{t} H_{D}(t) = \Omega \partial_{\varphi}H_{D}(\varphi)$ (where dimensionless variable $\varphi = \Omega t$) is much smaller than unity. It is known that in such a scenario, the dynamics is adiabatic \cite{garrison1988} and the system evolves through the instantaneous Bloch states $|\Psi_{\pm}(k,A(\varphi)) \rangle$, such that $H_{D}(\varphi) |\Psi_{\pm}(k,A(\varphi)) \rangle=E_{\pm}(k,A(\varphi))|\Psi_{\pm}(k,A(\varphi))\rangle$ where $|\Psi_{\pm}(k,A(\varphi)) \rangle=|\Psi_{\pm}(k+A(\varphi))\rangle$ and $E_{\pm}(k,A(\varphi))=E_{\pm}(k+A(\varphi))$. The corresponding cell periodic instantaneous Bloch states are then given by: $|u_{\pm}(k,A(\varphi)) \rangle = \mathcal{T}(-k)|\Psi_{\pm}(k,A(\varphi)) \rangle$, which solve the right eigenvalue problem: $H_{k + A(\varphi)} |u_{\pm}(k,A(\varphi)) \rangle = E_{\pm}(k,A(\varphi)) |u_{\pm}(k,A(\varphi)) \rangle$, where  
\begin{align}
H_{k + A(\varphi)} =   
\begin{pmatrix}
| 0 , A \rangle, & | 0 , B \rangle
\end{pmatrix}
[ H (k + A(\varphi))]
\begin{pmatrix}
\langle 0 , A| \\ \langle 0 , B|
\end{pmatrix},
\end{align} 
and $H(k + A(\varphi))$ reads:
\begin{align} \label{Hka} H(k + A(\varphi)) =
\begin{pmatrix}
0 & w e^{i (k + A(\varphi))} + v \\ w e^{\theta - i (k + A(\varphi))} + v  & 0
\end{pmatrix}.
\end{align}
From here it is evident that given an initial state with a fixed wavevector say $k_0$, the net effect of time evolution due to $A(\varphi)$ is same as changing $k_0$ to $k_0 + A(\varphi)$. Thus the state of the system at any instant of time is expressible as: $|u_{\pm}(k_0,A(\varphi)) \rangle \equiv |u_{\pm}(k_{0} + A(\varphi)) \rangle$ \cite{zak1989,pz2019}. As the parameter $\varphi$ changes, the state effectively sweeps the $k$-space so as to acquire the geometric phase $\gamma_{\pm}$ solely depending upon the ratio $\kappa$, as shown in the earlier section. Thus, we see that the driving potential, in this case, is unable to alter the system's topological structure as in the high-frequency case.


{\it Intermediate-frequency driving:} Finally, we here discuss the intermediate frequency of driving when $\Omega \approx v,w$. In this regime, the exact analytical solution of the eigenvalue problem in (\ref{fbproblem}) is intractable, and the approximate approaches employed in the earlier two frequency regimes fail. This is because the matrix $[\mathcal{H}_{m',m}(k)]$ as defined in (\ref{fbmatrix}) does not get simplified, since different Fourier blocks in $[\mathcal{H}_{m',m}(k)]$ get strongly coupled to each other in the intermediate-frequency regime. Nevertheless, we can numerically study the quasienergy spectrum of the driven SSH model in this regime applying PBC and OBC with $l=1/N$. 

\begin{figure}[h]
\includegraphics[width=8cm,height=10cm]{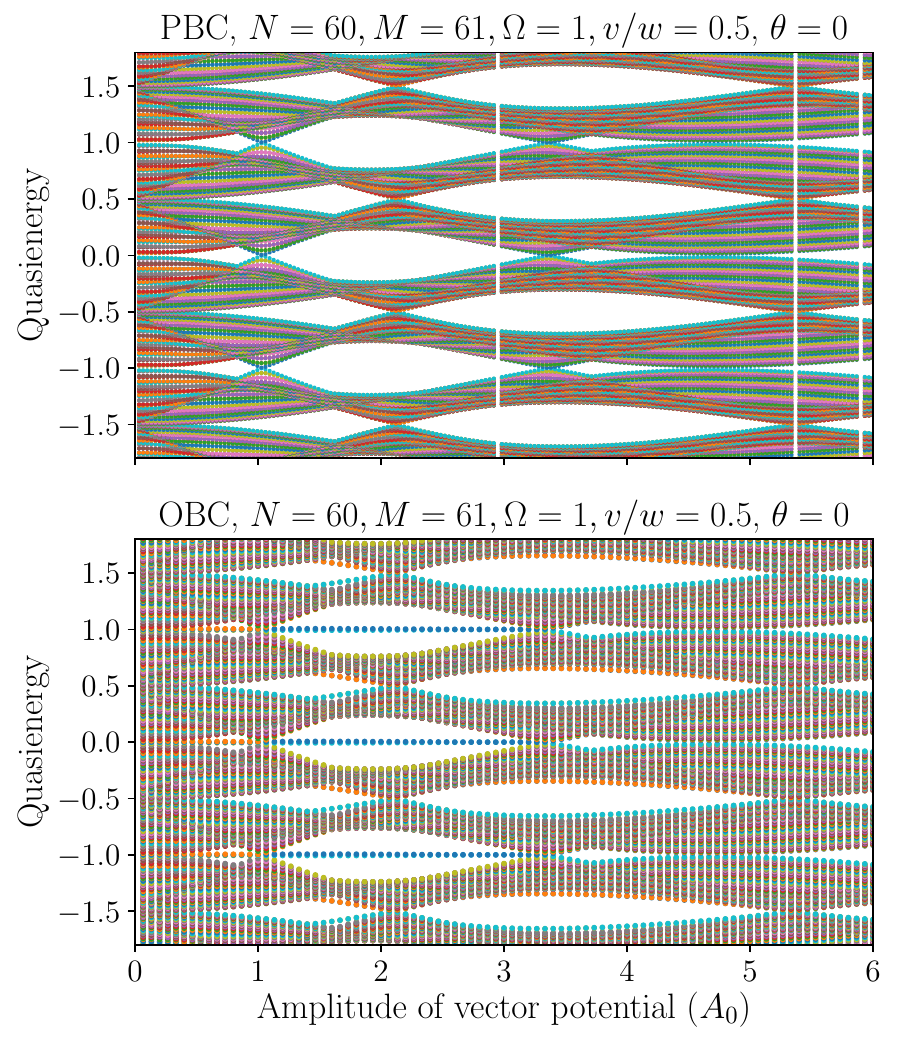}
\caption{Quasienergy spectrum versus amplitude of vector potential $(A_0)$ of a driven Hermitian SSH model at intermediate-frequency driving with PBC (top) and OBC (bottom). The two spectra closely match apart from the zero modes in the spectrum with OBC. All parameters are shown at the headings. 
  \label{quasiplot3}}
\end{figure}

\begin{figure}[h]
\includegraphics[width=8cm,height=10cm]{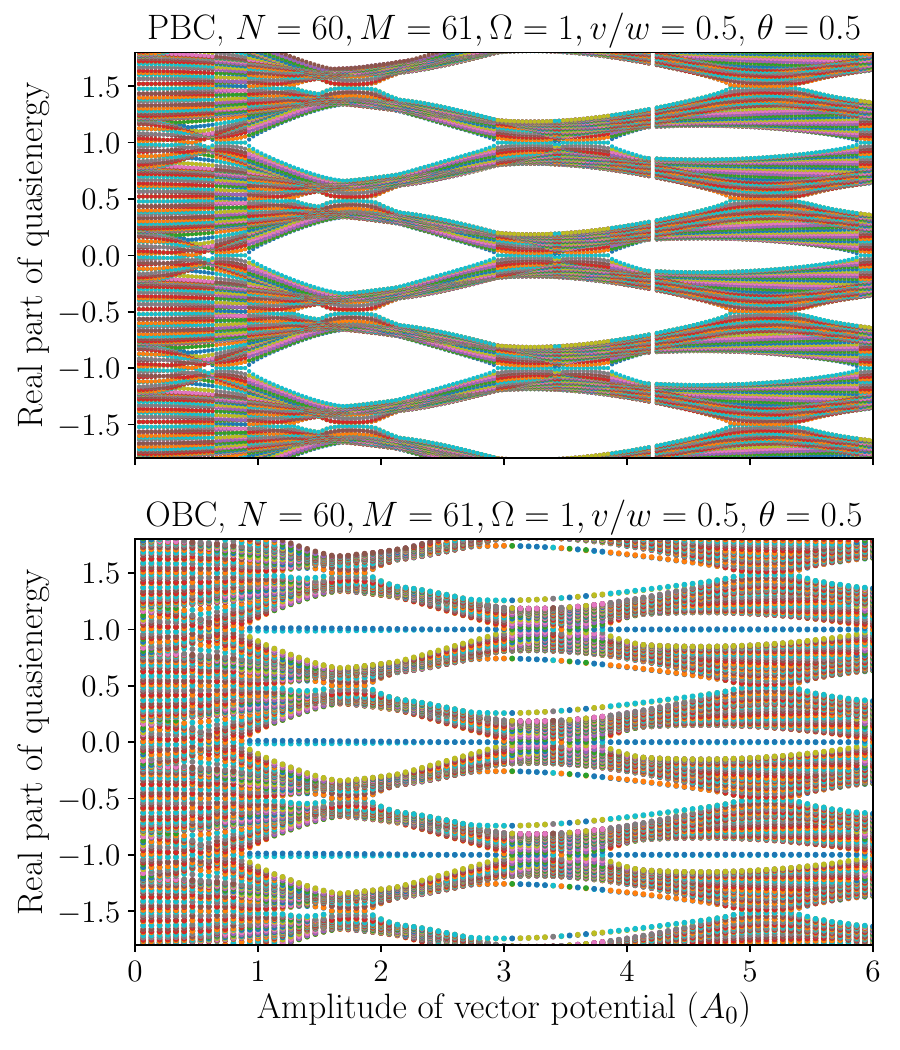}
\caption{Real part of quasienergy spectrum versus amplitude of vector potential $(A_0)$ of a driven non-Hermitian SSH model at intermediate-frequency driving with PBC (top) and OBC (bottom). The two spectra closely match, apart from the zero modes in the spectrum with OBC. While the size of the bulk band gap increases with a higher non-Hermiticity, the overlap of quasienergy bands indicating M{\"o}bius metallic phase is also clearly visible for the non-Hermitian model. All parameters are shown at the headings.\label{quasiplot4}} 
\end{figure}

In Fig.~ (\ref{quasiplot3}), we plot the quasienergy spectrum of the driven Hermitian SSH model for PBC and OBC. The two spectra closely match, apart from the zero modes in the spectrum with OBC. The quasienergy bands are periodic in energy with a period of $\Omega$, and two bands in each period (e.g., between $-\Omega/2$ and $\Omega/2$) touch each other near the topological phase transition as $A_0$ is varied. There is no M{\"o}bius metallic-like phase in the Hermitian model.   While the size of the bulk band gap increases with a higher non-Hermiticity in Fig.~(\ref{quasiplot4}), the overlap of quasienergy bands indicating the M{\"o}bius metallic-like phase is also clearly visible in the non-Hermitian model. The Fig.~(\ref{quasiplot4}) shows a gapped non-trivial topological phase with zero modes for OBC and a gapless  M{\"o}bius  phase within the range of $A_0$ in the figure. The gapped trivial phase appears at a higher value of $A_0$ for these parameters. Again, the two spectra in Fig.~(\ref{quasiplot4}) closely match apart from the zero modes in the spectrum with OBC. Nevertheless, a systematic characterization of different topological phases through a bulk topological invariant in the intermediate frequency of driving is yet to be carried out.

\section{Outlook}

In this paper, the topological aspects of a non-Hermitian generalization of the SSH model periodically driven by an external electromagnetic field are studied. A bi-orthonormal conception of the Pancharatnam-Zak geometric phase is constructed for the undriven model in the PBC framework. We have found that this geometric phase correctly captures the model's topological phase structure, which comprises trivial and non-trivial insulating phases and an exotic M{\"o}bius metallic phase. We also see the zero-energy modes when the model is investigated with OBC.  We note that implementing an OBC in this model requires care, and when implemented using the large $ N $ limit, consistent results are obtained.  We further discover that while the metallic phase is a topologically non-trivial phase, it does not support robust zero-energy modes, as displayed by the non-trivial insulating phase. 

Subsequently, the topological phase structure of the driven non-Hermitian SSH model is studied using the Floquet approach. It is found that this model also admits trivial and non-trivial insulator phases and the M{\"o}bius metallic phase. Interestingly, the driving potential's strength acts as a control parameter guiding the system through the various topological phases. We observe that the bi-orthogonal geometric phase corresponding to the  Floquet-Bloch states, in this case, acts as a topological index characterizing the different topological phases at high and low-frequency driving. Nevertheless, we could not resolve the nature of the topological phases at the intermediate-frequency driving. 

The present work unifies and extends the earlier works on Hermitian geometric phases due to Pancharatnam \cite{pancha1956}, Zak \cite{zak1989}, and the results on the non-Hermitian geometric phase of Garrison and Wright \cite{garrison1988} and others \cite{lieu2018}. Although several earlier studies define various topological invariants, which can capture and classify the topological phase structure of the driven non-Hermitian model, the bi-orthonormal geometric phase considered here is a straightforward elementary construction accomplishing this task. There have been earlier works dealing with topological aspects of driven non-Hermitian lattice models, wherein different kinds of non-Hermiticity and driving mechanisms were employed \cite{delplace2014,song2019,gong2018,torres2015,dalibard2014,kunst2018}.  We hope that the approach presented here will shed light on topological aspects in such models as well.

\begin*{\it Acknowledgments.}
We thank Prof. M. V. Berry for discussion at the initial stages of the work. DR further thanks Kiran Estake for pointing out an error in implementing PBC in Figs.~7,8 of the published version in the journal. This research is funded by the Department of Science and Technology, India via the Ramanujan Fellowship, and the Ministry of Electronics $\&$ Information Technology (MeitY), India under grant for ``Centre for Excellence in Quantum Technologies'' with Ref. No. 4(7)/2020-ITEA. 
\end*

\vskip1cm

\section*{Appendix}

For the general non-Hermitian Hamiltonian, wherein no assumption is made about the choice of parameters $v_{1,2}$ and $w_{1,2}$, it is straightforward to see that the determinant of $H$ (defined in (\ref{HSSH})) reads:
\begin{align}
\text{Det}\: H = (-1)^N(v_1^N +(-1)^{N-1}l w_2^N)(v_2^N +(-1)^{N-1} l w_1^N).
\end{align} 
The ratio $r$ for the general case is then given by:
\begin{align}
r = \frac{\text{Det}\: H \lvert_{OBC,N+1}}{\text{Det} \:H \lvert_{PBC,N}} \simeq - \frac{(v_1^N - \frac{1}{N} w_2^N)(v_2^N - \frac{1}{N} w_1^N)}{(v_1^N +w_2^N)(v_2^N + w_1^N)}.
\end{align}
Her,e we have assumed $N$ to be odd for specificity, and it can be checked that the end results are not affected by this assumption. Further analysis can be simplified by working with real parameters $\rho_{1,2}$ and $\phi_{1,2}$, defined as: $\frac{w_1}{v_2} = \rho_{1} e^{i \phi_{1}}$ and $\frac{w_2}{v_1} = \rho_{2} e^{i \phi_{2}}$. The ratio $r$ now reads:
\begin{align}
r \simeq - \frac{(1 - \frac{1}{N}  \rho_{1}^{N} e^{i N \phi_{1}})(1 - \frac{1}{N}  \rho_{2}^{N} e^{i N \phi_{2}})}{(1 +\rho_{1}^{N} e^{i N \phi_{1}})(1 +\rho_{2}^{N} e^{i N \phi_{2}})}.
\end{align}
This expression for $r$ immediately shows the phase diagram of this general model and the zero-mode energy scaling. It is clear that the system depending upon the Hamiltonian parameters exhibits three different phases classified as:
\begin{itemize}
\item[(a)] \underline{Trivial insulator phase}: when $\rho_{1} < 1$ and $\rho_{2} < 1$, no zero mode exists;

\item[(b)] \underline{Non-trivial insulator phase}: when $\rho_{1} > 1$ and $\rho_{2} > 1$, zero modes exist, and their energies fall as $\frac{1}{N}$;

\item[(b)] \underline{Metallic phase}: when $\rho_{1} < 1$ \&  $\rho_{2} > 1$ or when $\rho_{1} > 1$ \& $\rho_{2} < 1$, zero modes exist, and their energies fall as $\frac{i}{\sqrt{N}}$.
\end{itemize}

\bibliography{ref}	
\end{document}